\documentclass[12pt]{iopart}
\usepackage{iopams}  
\usepackage{graphicx}
\usepackage{bm}
\begin{document}
% Journal identifier can be put here if required, e.g.
%\jl{14}

\title{Non-Gaussianity of one-point distribution functions
in extended Lagrangian perturbation theory}

\author{Takayuki Tatekawa}

\address{\ Department of Physics, Waseda University,
3-4-1 Okubo, Shinjuku, Tokyo 169-8555, Japan}

\begin{abstract}
We study the one-point probability distribution functions (PDFs) of
the peculiar velocity and the density fluctuation in a cosmological
fluid. Within the perturbative approach to the structure formation
scenario, the effect of ``pressure'' has recently been
an area of research interest.
The velocity dispersion of the cosmological fluid creates
effective ``pressure'' or viscosity terms.
From this viewpoint, because the pressure reflects a nonlinear effect
of the motion of the fluid,
the pressure model would include nonlinear effects.
Here we analyze the Lagrangian linear perturbation PDFs
for both the Zel'dovich approximation and the pressure model.
We find that the PDFs of the peculiar velocity
remain Gaussian, even if we consider the pressure.
For the PDFs of the density fluctuation, the occurrence of
non-Gaussianity depends on the ``equation of state'' for the fluid.
Therefore we distinguish the ``equation of state''
using the PDFs.

\end{abstract}

%\submitted

\maketitle

%%%%%%%%%%%%%%%%%%%%%%%%%%%%%%%%%%%%%%%%%%%
\section{Introduction}\label{sec:intro}
%%%%%%%%%%%%%%%%%%%%%%%%%%%%%%%%%%%%%%%%%%%

The scenario for the formation of large-scale structure
in the Universe is based on the
gravitational instability of the primordial density fluctuation.
This fluctuation may have been generated from the
quantum fluctuations during the inflation
phase~\cite{Paddy93book,coles,Peacock}.
The fluctuation is spontaneously grown by
its self-gravitational instability.
To describe the evolution of the density fluctuation, many approaches
have been carried out.

For a perturbative approach,
the Lagrangian description provides
a relatively accurate model even in a quasi-linear stage.
Zel'dovich~\cite{zel} originally proposed a linear
Lagrangian approximation for dust (pressureless) fluid.
This approximation is called the Zel'dovich approximation
(ZA)~\cite{Paddy93book,coles,zel,Arnold82,Shandarin89,buchert89,saco,Jones,Tatekawa05,Paddy05}.
ZA describes the evolution of density fluctuation better than
the Eulerian
approximation~\cite{munshi,sahsha,Yoshisato98}. 
Although ZA gives an accurate description
until a quasi-linear regime develops,
ZA cannot describe the model after the formation
of caustics. In ZA, even after the formation of caustics, the fluid elements
keep moving in the direction set up by the initial condition.
In order to proceed with a hydrodynamical description
without the formation of caustics, although modified models,
such as the ``adhesion approximation''~\cite{gurbatov} and
the ``Truncated Zel'dovich approximation''
~\cite{cms,mps} were proposed, the
physical origin of the modification has not yet been clarified,
the physical origin of the modification has
not yet been clarified.

Recently, instead of the dust fluid, pressure-supported fluid has
been considered.
The collisionless Boltzmann equation~\cite{BT} describes the motion
of matter in phase space. The basic equations of hydrodynamics
are obtained by integrating the collisionless Boltzmann equation over
velocity space. 
In past approximations, such as ZA and its modified
models, velocity dispersion was ignored.
Buchert and Dom\'{\i}nguez~\cite{budo} argued that the effect
of the velocity dispersion should be noticeable beyond the caustics.
They showed that when the velocity dispersion
can be considered isotropic, it gives effective
``pressure'' or viscosity terms. Furthermore, they argued for
a relation between mass density $\rho$ and pressure $P$, i.e.,
an ``equation of state.'' If the relation between the density of matter
and pressure can be regarded as barotropic,
they showed that the ``equation of state'' should take the form
$P \propto \rho^{5/3}$.
Buchert {\it et al.}~\cite{bdp} showed how the viscosity term
is generated by the effective pressure of a fluid
under the assumption that the peculiar acceleration is
parallel to the peculiar velocity.
Dom\'{\i}nguez~\cite{domi00,domi02} introduced the idea that
the evolution equations for the matter fields are smoothed over
a smoothing length; then the viscosity term in
the ``adhesion approximation'' can be derived by the expansion
of coarse-grained equations. Recently, Buchert and
Dom\'{\i}nguez~\cite{Buchert05} discussed the origin of
the viscosity term and the extension of the Lagrangian
perturbation theory.

Departing from these points of view,
the extension of the Lagrangian perturbation
theory to cosmological fluids with pressure has been considered.
Adler and Buchert~\cite{adler} formulated the 
Lagrangian perturbation theory for a barotropic fluid.
Morita and Tatekawa~\cite{Morita01} and
Tatekawa {\it et al.}~\cite{Tatekawa02}
solved the Lagrangian perturbation equations for a polytropic fluid
up to the second order. Recently, Tatekawa~\cite{Tatekawa05A} solved
these same equations up to the third order.
Hereafter, we call this model the ``pressure model.''
Buchert and Dom\'{\i}nguez~\cite{Buchert05} call this model
the ``Euler-Jeans-Newton'' (EJN) model.

In this paper, we study the occurrence of non-Gaussianity in the
one-point probability density functions (PDFs). For one-point PDFs
of the smoothed density and velocity fields in a cosmological
model, Kofman {\it et al.}~\cite{Kofman94} analyzed
the evolution beginning with a Gaussian initial fluctuation.
Their analytic results are based on ZA. They found that the
PDF of the peculiar velocity, both Eulerian and Lagrangian,
remains Gaussian under linear approximation. For the density
fluctuation, they showed that the PDF develops a shape similar
to a lognormal distribution.

We consider PDFs as follows. From the point at which
the pressure is derived from the collisionless Boltzmann
equation (for example, \cite{budo, bdp, domi00,domi02}),
even if we consider only linear perturbation,
the pressure model would include nonlinear effects. Therefore,
even if we analyze the evolution of the peculiar velocity PDF
only with linear approximation, it will deviate from Gaussian.
Please note that the contribution of
the pressure can be distinguished with PDFs. 
The effect of the pressure changes the shape and the evolution
of the PDFs.

We analyze non-Gaussianity of the PDF of the peculiar velocity
and density
fluctuation. For our analyses, we compute skewness and kurtosis,
which are statistical quantities for non-Gaussianity of the PDF. 
First, we show the results for the dust model with ZA
and N-body simulation.
Our results coincide with those of previous efforts.
Then we indicate
that the skewness and the kurtosis are useful for the analysis of
non-Gaussianity. Next we analyze the PDFs for the pressure model.
Here we consider both the Eulerian and the Lagrangian linear
approximations. In the pressure model, because the growing
rate of the fluctuation depends on the scale, the occurrence
of the non-Gaussianity was expected. However, under linear
approximation,
the effect of the pressure does not contribute to occurrence of
the non-Gaussianity very much. The PDF of the peculiar velocity
remains largely Gaussian during evolution. For the PDF of the
density fluctuation, the Eulerian linear approximation does
not indicate the occurrence of non-Gaussianity. On the other hand,
in the Lagrangian linear approximation non-Gaussianity
does obviously occur. Furthermore, the evolution of the PDF
depends on the ``equation of state.'' Therefore if we
analyze carefully the PDFs obtained from observations,
we can expect to find
a constraint to the ``equation of state'' for the cosmological fluid.

This paper is organized as follows.
In Sec.~\ref{sec:equation}, we briefly present the evolution equation
for cosmological fluid. For simplicity, we consider only the Einstein-de
Sitter Universe model.

In Sec.~\ref{sec:NonGauss}, we analyze non-Gaussianity of the one-point
PDFs. Here we introduce two statistical quantities, skewness and kurtosis.
First, in Sec.~\ref{subsec:NG-dust}, we analyze the dust model.
According to past study, the PDF of the density fluctuation
develops a shape similar to a lognormal distribution. Its
non-Gaussianity could be detected with the skewness and kurtosis.
In Sec.~\ref{subsec:NG-P}, we analyze the pressure model.
In the dust model, the PDFs in the Eulerian linear approximation
remains Gaussian. Also in the pressure model, this tendency was
unchanged. The PDF of the density fluctuation in the
Lagrangian linear approximation was sensitive to the
``equation of state.''
 
In Sec.~\ref{sec:discuss}, we discuss our results
and state conclusions.

%%%%%%%%%%%%%%%%%%%%%%%%%%%%%%%%%%%%%%%%%%%
\section{The evolution of fluctuation for the cosmological fluid}
\label{sec:equation}
%%%%%%%%%%%%%%%%%%%%%%%%%%%%%%%%%%%%%%%%%%%

Here we briefly introduce the evolution equation for cosmological
fluid. In the comoving coordinates,
the basic equations for cosmological fluid are described as
\begin{eqnarray}
\frac{\partial \delta}{\partial t} + \frac{1}{a} \nabla_x \cdot \{ \bm{v}
(1+\delta) \} &=& 0 \,, \label{eqn:comoving-conti-eq} \\
\frac{\partial \bm{v}}{\partial t} + \frac{1}{a} (\bm{v} \cdot \nabla_x)
\bm{v} + \frac{\dot{a}}{a} \bm{v} &=& \frac{1}{a} \tilde{\bm{g}} - \frac
{1}{a \rho} \nabla_x P \,, \label{eqn:comoving-Euler-eq}
\end{eqnarray}
\begin{eqnarray}
\nabla_x \times \tilde{\bm{g}} &=& \bm{0} \,, \label{eqn:rot-g} \\
\nabla_x \cdot \tilde{\bm{g}} &=& - 4 \pi G \rho_b a \delta \,,
\label{eqn:comoving-Poisson-eq} \\
\delta & \equiv & \frac{\rho - \rho_b}{\rho_b} \,.
\end{eqnarray}
In the Eulerian perturbation theory, the density fluctuation $\delta$
is regarded as a perturbative quantity. In linear approximation,
from equation~(\ref{eqn:comoving-conti-eq}), we obtain
\begin{equation} \label{eqn:EA-v-delta}
\frac{\partial \delta}{\partial t} + \frac{1}{a} \nabla_x \cdot \bm{v}
= 0 \,.
\end{equation}
Then we take a divergence of equation~(\ref{eqn:comoving-Euler-eq}) and
substitute equation~(\ref{eqn:EA-v-delta}) to it. Finally
we obtain the evolution equation for the density fluctuation
in the Eulerian linear approximation~\cite{Weinberg}:
\begin{equation} \label{eqn:EA-lin-delta}
\frac{\partial^2 \delta}{\partial t^2} + 2 \frac{\dot{a}}{a}
 \frac{\partial \delta}{\partial t} - 4 \pi G \rho_b \delta
 - \frac{1}{\rho_b a} \nabla_x^2 P =0 \,.
\end{equation}
When we assume a polytropic fluid ($P = \kappa \rho^{\gamma}$),
equation~(\ref{eqn:EA-lin-delta}) becomes
\begin{equation} \label{eqn:EA-lin-delta2}
\frac{\partial^2 \delta}{\partial t^2} + 2 \frac{\dot{a}}{a}
 \frac{\partial \delta}{\partial t} - 4 \pi G \rho_b \delta
 - \frac{\kappa \gamma \rho_b^{\gamma-1}}{a^2} \nabla_x^2 \delta =0 \,.
\end{equation}
On the other hand, in the
Lagrangian perturbation theory, the displacement from homogeneous
distribution is considered.
\begin{equation} \label{eqn:x=q+s}
\bm{x} = \bm{q} + \bm{s} (\bm{q},t) \,,
\end{equation}
where $\bm{x}$ and $\bm{q}$ are the comoving Eulerian coordinates
and the Lagrangian coordinates, respectively. $\bm{s}$ is
the displacement vector that is regarded as a perturbative quantity.
From equation~(\ref{eqn:x=q+s}), we can solve the continuous
equation~(\ref{eqn:comoving-conti-eq}) exactly. Then the density
fluctuation is given in the formally exact form.
\begin{equation} \label{eqn:L-exactrho}
\delta = 1 - J^{-1}, ~~ J \equiv \det \left (
 \frac{\partial x_i}{\partial q_j} \right ) \,.
\end{equation}
$J$ means the Jacobian of the coordinate transformation
from Eulerian $\bm{x}$ to Lagrangian $\bm{q}$.
Therefore when we derive the solutions for $\bm{s}$, we can know
the evolution of the density fluctuation. 

The peculiar velocity is given by
\begin{equation}
\bm{v}=a \dot{\bm{s}} \label{eqn:L-velocity} \,.
\end{equation}
To solve the perturbative equations,
we decompose the Lagrangian perturbation to the longitudinal
and the transverse modes:
\begin{eqnarray}
\bm{s} &=& \nabla S + \bm{s}^T \,, \\
\nabla \cdot \bm{s}^T &=& 0 \,,
\end{eqnarray}
where $\nabla$ means the Lagrangian spacial derivative.
Hereafter we consider only the longitudinal mode.

Using the Lagrangian displacement, we obtain the first-order
perturbative equation.
\begin{equation} \label{eqn:LA-lin}
\nabla^2 \left ( \ddot{S}^{(1)} + 2 \frac{\dot{a}}{a}
 \dot{S}^{(1)} - 4\pi G \rho_b S^{(1)}
  - \frac{\kappa \gamma \rho_b^{\gamma-1}}{a^2} \nabla^2 S^{(1)}
 \right ) = 0 \,.
\end{equation}
Let us compare equations~(\ref{eqn:EA-lin-delta2}) and (\ref{eqn:LA-lin}).
In both of the equations, the same operator acts on the perturbative
quantity~\cite{Morita01}.
Therefore in the linear approximation, the form of the 
Lagrangian solutions is the same as that of the Eulerian solutions.

The first-order solutions for the longitudinal mode depend on
spacial scale. Therefore the solutions are described with a
Lagrangian wavenumber ${\bm{K}}$. For simplicity, in this paper,
we discuss only perturbative
solutions in the Einstein-de Sitter Universe
model~\cite{Morita01, Tatekawa02}.
\begin{eqnarray}
\widehat{S}^{(1)} (\bm{K}, t) &=& C^+ (\bm{K}) D^+ (\bm{K}, t)
 + C^- (\bm{K}) D^- (\bm{K}, t) \,, \\
D^{\pm} (\bm{K}, t) &=& 
\left \{
\begin{array}{lcl}
t^{-1/6} {\cal J}_{\pm 5/(8-6 \gamma)} (A | \bm{K} | t^{-\gamma+4/3})
& \mbox{for} & \gamma \ne \frac{4}{3} \,, \\
t^{-1/6 \pm \sqrt{25/36 - B|\bm{K}|^2}}
& \mbox{for} & \gamma  = \frac{4}{3} \,, \\
\end{array}
\right.
\end{eqnarray}
\[
A \equiv \frac{3 \sqrt{\kappa \gamma} (6 \pi G)^{(1-\gamma)/2}}{|4-3\gamma|},
~ B \equiv \frac{4}{3} \kappa (6 \pi G)^{-1/3} \,,
\]
where ${\cal J}$ means Bessel function.
$C^{\pm} (\bm{K})$ is given by the initial condition.

In this model, the behavior of the solutions strongly depends on
the relation between the scale of fluctuation and the Jeans scale.
Here we define the comoving Jeans wavenumber as
\begin{equation} \label{eqn:Pressure-Jeans}
K_{\rm J} \equiv \left(
\frac{4\pi G a^2}
     {\kappa \gamma \rho_{\rm b}^{\gamma-2}} \right)^{1/2} \,.
\end{equation}
It depends on the ratio between the comoving Jeans wavenumbers
and the wavenumber of the fluctuation whether the fluctuation
grows.
When we ignore the pressure term, we obtain the solutions of ZA.
\begin{equation}
S^{(1)} = t^{2/3} S_+ (\bm{q}) + t^{-1} S_- (\bm{q}) \,.
\end{equation}

$S_{\pm} (\bm{q})$ is given by the initial condition.

%%%%%%%%%%%%%%%%%%%%%%%%%%%%%%%%%%%%%%%%%%%
\section{Non-Gaussianity of the one-point distribution functions}
\label{sec:NonGauss}
%%%%%%%%%%%%%%%%%%%%%%%%%%%%%%%%%%%%%%%%%%%

We analyze the non-Gaussianity of the PDF of the peculiar velocity
and the density fluctuation. To investigate non-Gaussianity,
we compute two statistical quantities~\cite{Peacock,Kofman94,Peebles80}.
\begin{eqnarray*}
\mbox{skewness} &:& \left < \left (\frac{p- \left <p \right >}{\sigma_p}
 \right )^3 \right > \,, \\
\mbox{kurtosis} &:& \left < \left (\frac{p- \left <p \right >}{\sigma_p}
 \right )^4 \right > - 3 \,,
 \\
&& \sigma_p \equiv \left < \left (p- \left <p \right > \right )^2
 \right > , \mbox{$p$: physical quantity} \,.
\end{eqnarray*}
The skewness and the kurtosis shows display asymmetry and a non-Gaussian
degree of ``peakiness.''
If the distribution is completely Gaussian, both the skewness
and the kurtosis become zero. 

For simplicity,  we set the Gaussian density field with
a scale-free spectrum:
\begin{equation}
\mathcal{P}(k) \propto k \,.
\end{equation}
To avoid divergence of the density fluctuation, we introduce a
top-hat cutoff at $k^{-1} = 1 h^{-1} \mbox{Mpc}$ (at $a=1$) in comoving Eulerian coordinates.

The Gaussian initial condition is generated by COSMICS~\cite{COSMICS}.
We set up the initial condition at $a=10^{-3}$.
The initial peculiar velocity and the density fluctuation are adjusted
by the growing solution of ZA.
Both the skewness and the kurtosis
of the initial condition are less than $10^{-2}$.

For N-body simulation, we execute $P^3 M$ code. The parameters
of simulation were given as follows:
\begin{eqnarray*}
\mbox{Number of particles} &:& N=128^3 \,, \\
\mbox{Box size} &:& L=128 h^{-1} \mbox{Mpc}~~ (\mbox{at}~a=1) \,, \\
\mbox{Softening Length} &:& \varepsilon= 0.05 h^{-1} \mbox{Mpc}
~~(\mbox{at}~a=1)  \,.
\end{eqnarray*}

For perturbative models, we set the parameters as follows:
\begin{eqnarray*}
\mbox{Number of grids} &:& N=128^3 \,, \\
\mbox{Box size} &:& L=128 h^{-1} \mbox{Mpc}
 ~~(\mbox{at}~a=1)  \,.
\end{eqnarray*}

The dispersion, skewness and kurtosis is computed by
numerical method.
For analyses of non-Gaussianity, we coarsen the fields.
For the density fluctuation, we coarsen the density field over
$1 h^{-1} \mbox{Mpc}$ (at $a=1$) in comoving Eulerian coordinates
with a top-hat window function.
For the peculiar velocity, we analyze on grids
at intervals of $1 h^{-1} \mbox{Mpc}$ (at $a=1$)
in Lagrangian space
for Lagrangian perturbation models. In N-body simulation,
we analyze the peculiar velocity of each particle.

We compute the dispersion, skewness and kurtosis
at ten time slices from $a=0.1$ to $a=1.0$. The time interval
is given as $\Delta a=0.1$.

%%%%%%%%%%%%%%%%%%%%%%%%%%%%%%%%%%%%%%%%%%%
\subsection{Non-Gaussianity in the dust model}
\label{subsec:NG-dust}
%%%%%%%%%%%%%%%%%%%%%%%%%%%%%%%%%%%%%%%%%%%

First, we analyze the dust model. Here we analyze the PDF of
the peculiar velocity
and the density fluctuation for the Eulerian
linear approximation, ZA and N-body simulation.
For the dust models, Kofman {\it et al.}~\cite{Kofman94}
analyzed one point PDFs in detail.
They proved that for the PDF of the peculiar velocity, the Lagrangian
PDF is equal to the Eulerain PDF at all times.
For the PDF of the density fluctuation, Padmanabhan and
Subramanian~\cite{Paddy93} derived the distribution function
in the nonlinear regime using ZA. 

\begin{figure}[tb]
\centerline{
\includegraphics[height=7cm]{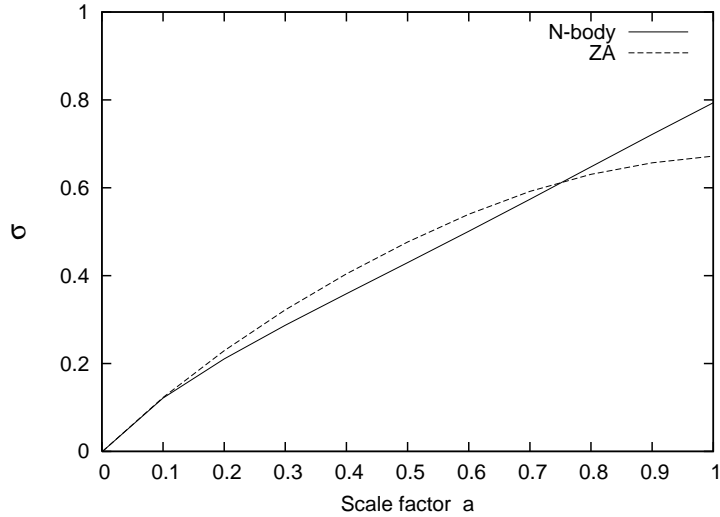}
}
\caption{The dispersion of the density fluctuation in ZA and
N-body simulation. Because of shell-crossing, the growth of
the density fluctuation becomes slow gradually.
}
\label{fig:dust-sigma}
\end{figure}

Now we analyze the quasi-nonlinear stage. Figure~\ref{fig:dust-sigma}
shows the dispersion of the density fluctuation. In our model,
the dispersion becomes $\sigma >0.1$ at $a=0.1$. Because
of shell-crossing, the growth of the density fluctuation
becomes slow gradually.

\begin{figure}[tb]
\centerline{
\includegraphics[height=7cm]{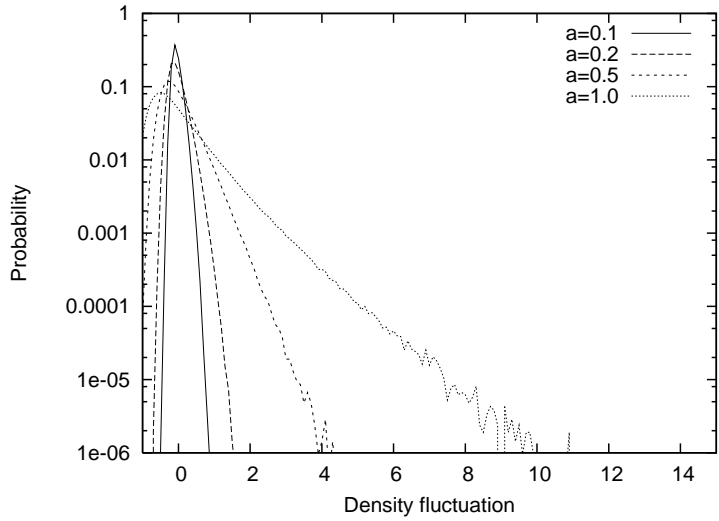}
}
\caption{The PDF of the density fluctuation in N-body simulation.
During evolution, the PDF approaches lognormal distribution.
}
\label{fig:Prob-delta}
\end{figure}

\begin{figure}[tb]
\centerline{
\includegraphics[height=7cm]{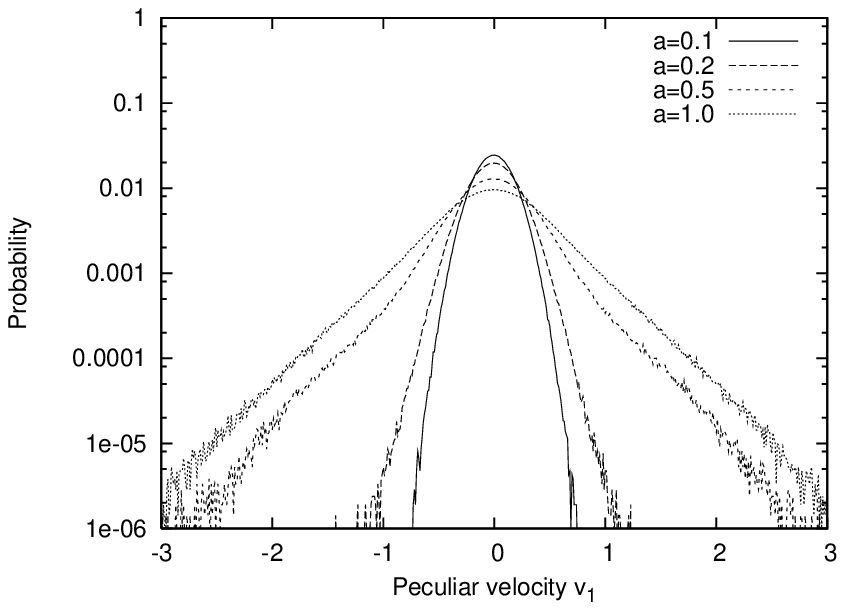}
}
\caption{The PDF of the peculiar velocity in N-body simulation.
During evolution, because of attraction to high density region,
the PDF deviates from Gaussian distribution.
}
\label{fig:Prob-V}
\end{figure}

Figure~\ref{fig:Prob-delta} shows the PDF of the density fluctuation
in N-body simulation. During evolution, the PDF approaches
log-normal distribution. The result coincides with that of
the past
analyses~\cite{Kofman94, Paddy93, Kayo01}.
Figure~\ref{fig:Prob-V} shows the PDF of the peculiar
velocity.
Because we set up almost isotropic initial condition,
even if we pay attention to the one direction, we
think that the generality of the result is not lost.
Therefore, we choose the x-direction peculiar velocity.
During evolution, because the dense region attracts
surrounding matter, the probability of fast velocity increases.
At $a=1.0$, both the density fluctuation and the peculiar
velocity obviously show non-Gaussian distribution.
Therefore the skewness and the kurtosis must have non-zero
value.

\begin{figure}[tb]
\centerline{
\includegraphics[height=7cm]{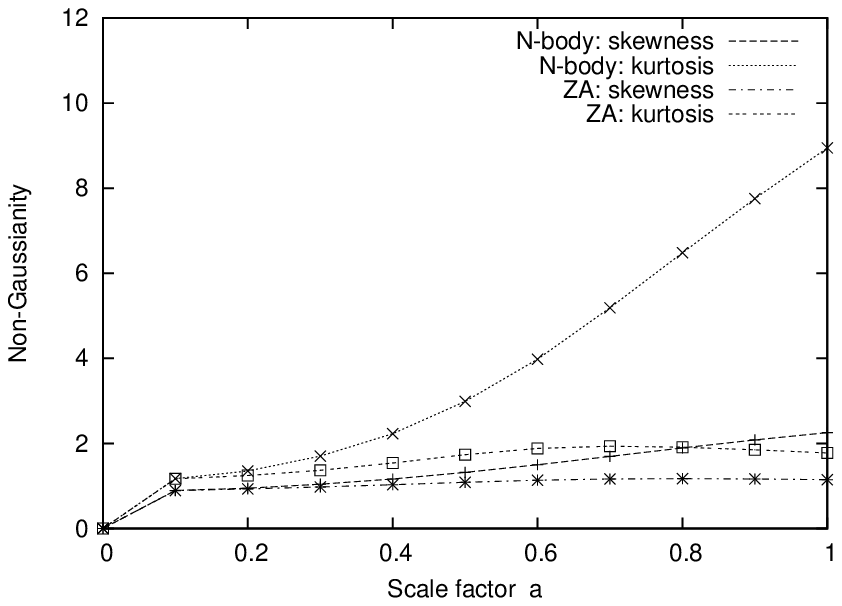}
}
\caption{The skewness and the kurtosis of the density fluctuation.
After quasi-nonlinear stage ($\delta_{\mbox{max}} >1$),
the positive fluctuation evolves quickly. Then the skewness increases
remarkably. Of course N-body simulation realizes strongly
nonlinear evolution; the skewness and the kurtosis increase
rapidly. For ZA, because of shell-crossing, the evolution
of the skewness and the kurtosis stops gently.
}
\label{fig:dust-delta-NG}
\end{figure}

Figure~\ref{fig:dust-delta-NG} shows the skewness and the kurtosis
of the density fluctuation. 
In N-body simulation, the PDF of
the density fluctuation is well-fitted by log-normal distribution
\cite{Kofman94}. Therefore N-body simulation realizes strongly
nonlinear evolution, then the skewness and the kurtosis increases
rapidly.
In ZA, the skewness and the kurtosis also increases during
evolution. However, ZA can describe only quasi-nonlinear evolution,
and after shell-crossing, i.e., the formation of the caustic,
ZA cannot describe the evolution of nonlinear structure.
Therefore the evolution of the skewness and the kurtosis
stops gently.

%%% Figure %%%
%%%%%%%%%%%%%%

%%% Figure %%%
%%%%%%%%%%%%%%

%%% Figure %%%
%%%%%%%%%%%%%%

\begin{figure}[tb]
\centerline{
\includegraphics[height=7cm]{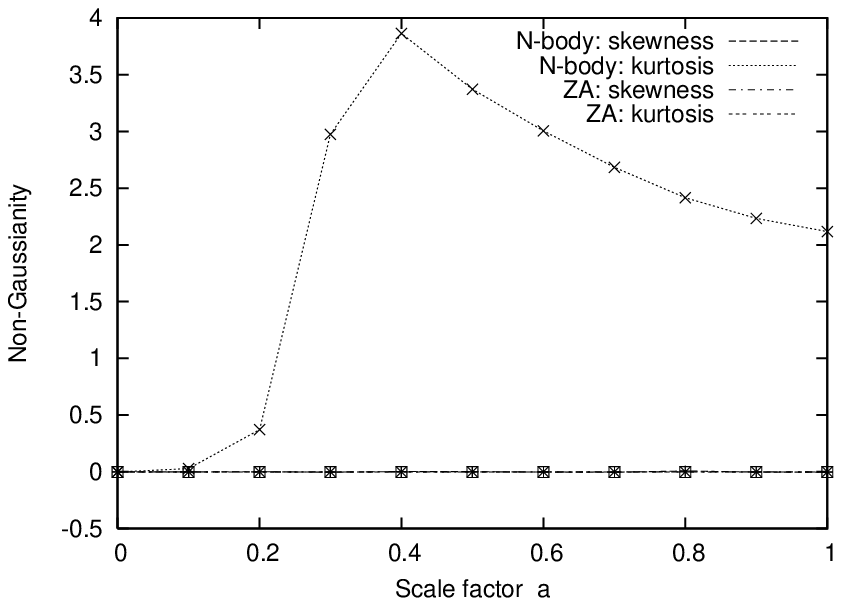}
}
\caption{The skewness and the kurtosis of the peculiar velocity.
Here we choose one direction of the peculiar velocity.
In N-body simulation, the high-density region attracts
surrounding matter, and the kurtosis increases. On the other hand,
when we set an isotropic initial condition, the skewness almost does
not increase. In ZA, because the evolution of the peculiar
velocity is always linear, the PDF of the peculiar velocity
remains Gaussian at all times.}
\label{fig:dust-V-NG}
\end{figure}

Figure~\ref{fig:dust-V-NG} shows the skewness and the kurtosis
of the peculiar velocity. Because we set initial condition by
almost isotropic distribution, the initial PDF of one direction
of the peculiar velocity is Gaussian. 
As a result the skewness and the kurtosis increase rapidly.
In ZA, the skewness and the kurtosis also increase during
evolution. However, ZA can describe only quasi-nonlinear evolution,
and after shell-crossing, i.e., the formation of the caustic,
ZA cannot describe the evolution of nonlinear structure.
Therefore the evolution of the skewness and the kurtosis
stops gently.
In the Eulerian linear approximation, the PDF of the peculiar velocity
and the density fluctuation remains Gaussian at all times.

%%% Figure %%%
%%%%%%%%%%%%%%

%%% Figure %%%
%%%%%%%%%%%%%%

From our analyses of the PDF in the dust model, we can conclude that
both the skewness and the kurtosis are useful quantities
for analyses of non-Gaussianity. Next we adopt these analyses
for the pressure model.

%%%%%%%%%%%%%%%%%%%%%%%%%%%%%%%%%%%%%%%%%%%
\subsection{Non-Gaussianity in the pressure model}
\label{subsec:NG-P}
%%%%%%%%%%%%%%%%%%%%%%%%%%%%%%%%%%%%%%%%%%%

In this subsection, we analyze the pressure model. Here we choose
the polytropic index $\gamma=4/3, 5/3$. In a previous
paper~\cite{Tatekawa04A},
we compared the density field between N-body simulation and
the Lagrangian approximations. In this comparison, we showed
that if a small value was chose for the initial Jeans wavelength
$K_J$ (equation~(\ref{eqn:Pressure-Jeans})),
it is hard for a nonlinear structure to form 
the pressure suppress the evolution of the density
fluctuation.
On the other hand, if we choose a large $K_J$ value, the
evolution of the density fluctuation becomes almost same as
that in the dust model. Therefore if we want to consider the effect
of the pressure and the formation of the nonlinear structure,
we should choose an appropriate value for $K_J$. In this paper, following
the results of our previous paper~\cite{Tatekawa04A}, we choose that
the  initial Jeans wavelength $K_J = 64$ at $a=10^{-3}$.
Here we do not specify
a dark matter model; we just analyze the behavior of the 
pressure models.
In the pressure model, the evolution of the perturbation depends
on the scale. Therefore even if the initial distribution is
Gaussian, we expect that the non-Gaussianity would appear
during evolution.

\begin{figure}[tb]
\centerline{
\includegraphics[height=7cm]{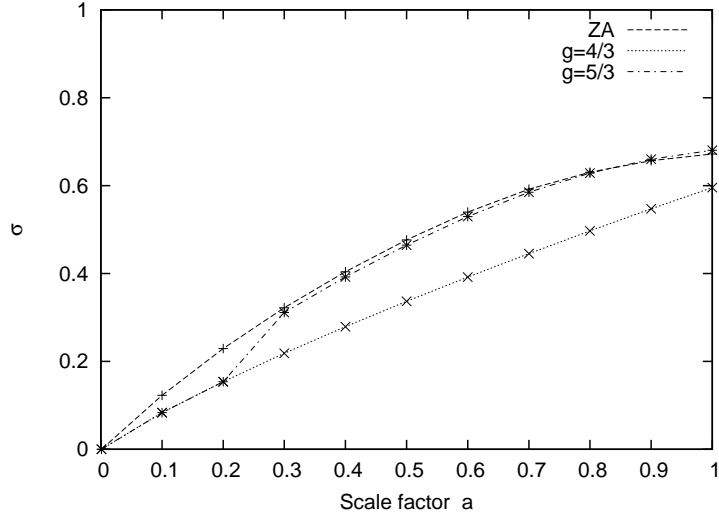}
}
\caption{The dispersion of the density fluctuation in ZA and
the pressure models. During evolution, the dispersion in the
case of $\gamma=5/3$ approaches to that in ZA.
}
\label{fig:P-sigma}
\end{figure}

Figure~\ref{fig:P-sigma} shows the dispersion of the density fluctuation.
In past papers~\cite{Morita01,Tatekawa02}, we show that 
the asymptotic behavior of the perturbative solution in the case
of $\gamma=5/3$ is similar to that in ZA. Here we show that the
statistical quantities in the case of $\gamma=5/3$ approach
to these in ZA during evolution. As was done with the dust model,
we analyze the quasi-nonlinear stage in the pressure model.

%%% Figure %%%
%%%%%%%%%%%%%%

%%% Figure %%%
\begin{figure}[tb]
\centerline{
\includegraphics[height=7cm]{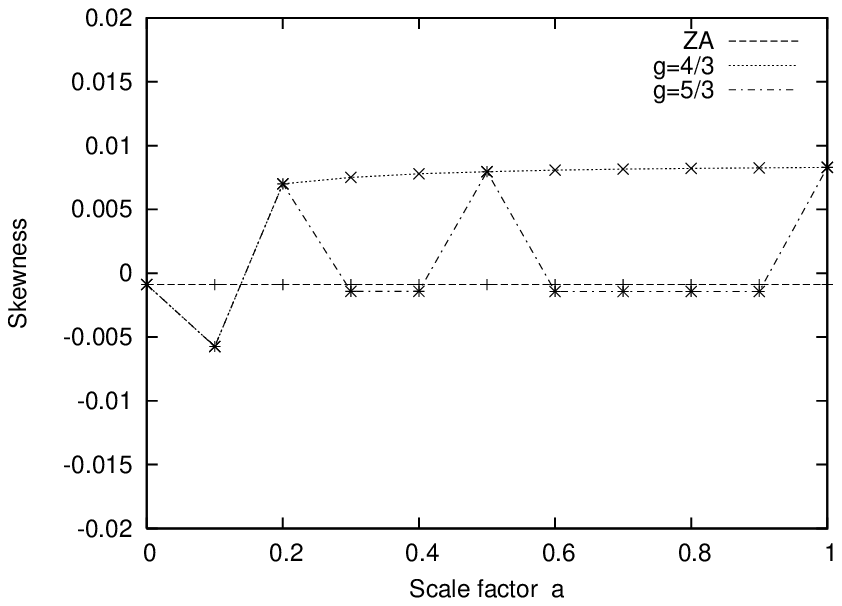}
}
\caption{The skewness
of the peculiar velocity in the pressure model.
Here we choose one direction of the peculiar velocity.
In the dust model, because the evolution of the peculiar
velocity is always linear, the skewness
takes a constant value.
In the pressure model, although the non-Gaussianity appears
during evolution, its extent remains small
in the quasi-nonlinear stage.}
\label{fig:P-V-m3}
\end{figure}
%%%%%%%%%%%%%%

%%% Figure %%%
\begin{figure}[tb]
\centerline{
\includegraphics[height=7cm]{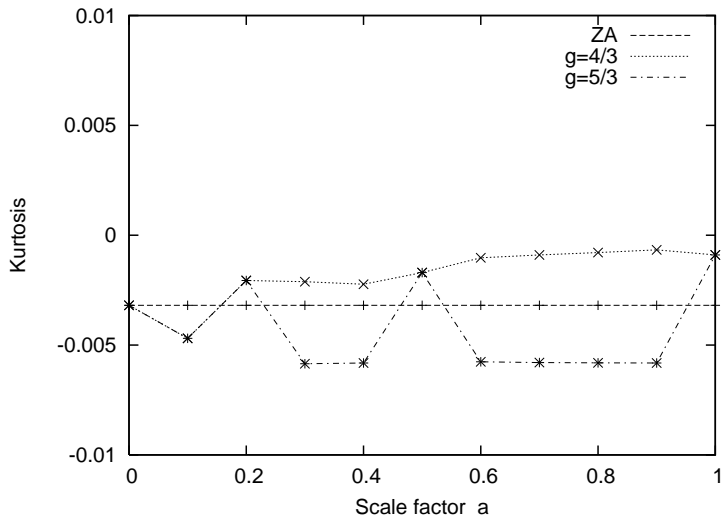}
}
\caption{The kurtosis of the peculiar velocity
in the pressure model.
Here we choose one direction of the peculiar velocity.
As in the case of the skewness,
in the dust model, because the evolution of the peculiar
velocity is always linear, the kurtosis
takes a constant value. 
In the pressure model, although the non-Gaussianity appears
during evolution, its extent remains small
in the quasi-nonlinear stage.}
\label{fig:P-V-m4}
\end{figure}
%%%%%%%%%%%%%%

Figures~\ref{fig:P-V-m3} and \ref{fig:P-V-m4}
shows the skewness and the kurtosis
of the peculiar velocity in the pressure model, respectively.
Under the linear perturbation, for the peculiar velocity,
the Lagrangian PDF is equal to the Eulerain PDF at all times
\cite{Kofman94}.
In the pressure model, although the non-Gaussianity appears
during evolution, its extent remains small
in the quasi-nonlinear stage. In the dust model, the PDF of the
peculiar velocity never changes. Even if
the nonlinear structure forms, the motion of the matter
does not stop. On the other hand, because the pressure
especially affects the dense region, the motion of the matter
is slowed down. Therefore, as we show in Figure~\ref{fig:P-V-m3},
the skewness of the peculiar velocity in the pressure model
grows more than ten times bigger than that in ZA. 

\begin{figure}[tb]
\centerline{
\includegraphics[height=7cm]{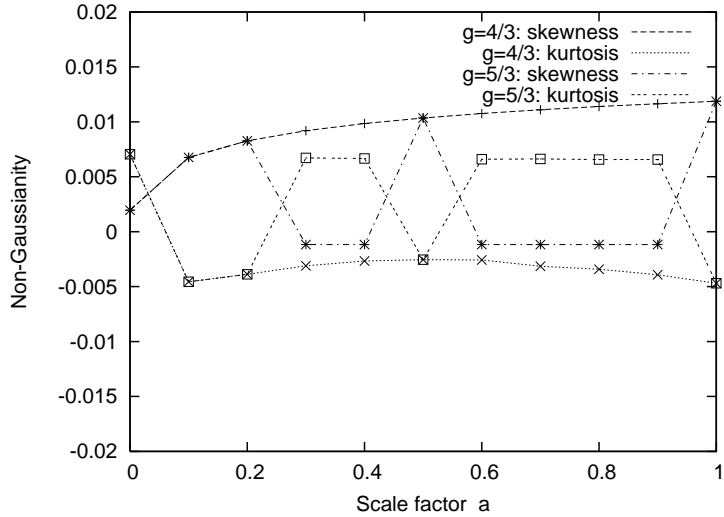}
}
\caption{The skewness and the kurtosis
of the density fluctuation in the Eulerian models.
Although we consider the effect of the pressure,
the PDF of the density fluctuation is almost Gaussian
during evolution.}
\label{fig:P-delta-EA-NG}
\end{figure}

Next we show the skewness and the kurtosis of the
density fluctuation in the
pressure model. First we analyze the Eulerian
linear approximation.
Even if we consider the Eulerian linear approximation,
the evolution of the fluctuation is not linear,
because of the pressure. The evolution
depends on the scale of the fluctuation. Therefore the
PDF of the density fluctuation may deviate from Gaussian.
Figure~\ref{fig:P-delta-EA-NG} shows the result. Although
the skewness and the kurtosis oscillate, the value is still
small in a nonlinear regime. Therefore we conclude that
the PDF of the density fluctuation almost retains Gaussianity.

%%% Figure %%%
%%%%%%%%%%%%%%

We analyze the PDFs in the Lagrangian linear approximation.
As we show in equation~(\ref{eqn:L-exactrho}), the relation
between the density fluctuation and
the Lagrangian displacement is nonlinear.
Therefore as we show
in the case of the dust model, even if we consider
the linear approximation, the PDF of the density
fluctuation deviates from Gaussian.

%%% Figure %%%
\begin{figure}[tb]
\centerline{
\includegraphics[height=7cm]{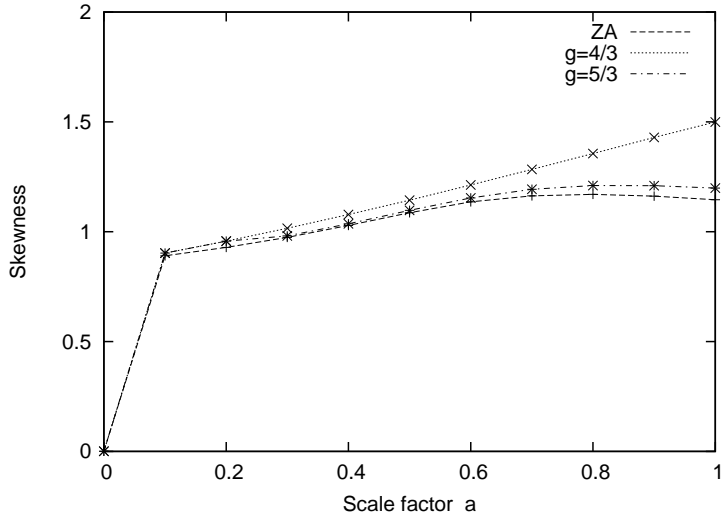}
}
\caption{The skewness
of the density fluctuation in the Lagrangian models.
In the dust model, because of shell-crossing,
the growth of the skewness stops gently.
In the pressure model, although the growth of the skewness
lasts longer than in the case of the dust model,
it finally stops because of shell-crossing.}
\label{fig:P-delta-m3}
\end{figure}
%%%%%%%%%%%%%%

%%% Figure %%%
\begin{figure}[tb]
\centerline{
\includegraphics[height=7cm]{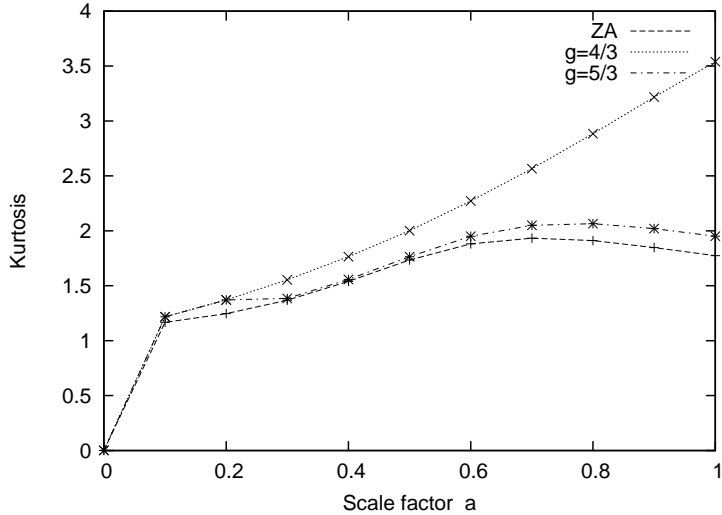}
}
\caption{The kurtosis of the density fluctuation
in the Lagrangian models.
As in the case of the skewness,
in the dust model, because of shell-crossing,
the growth of the kurtosis stops gently. 
In the pressure model, although the growth of the kurtosis
lasts longer than in the case of the dust model,
it finally stops because of shell-crossing.}
\label{fig:P-delta-m4}
\end{figure}
%%%%%%%%%%%%%%

Figures~\ref{fig:P-delta-m3} and \ref{fig:P-delta-m4}
show the skewness and the kurtosis of the density fluctuation.
In the dust model, because of shell-crossing,
the growth of the skewness stops gently.
In the pressure model, although the growth of the skewness
lasts longer than in the case of the dust model,
it finally stops because of shell-crossing.
The behavior of the growth of the non-Gaussianity
depends on the ``equation of state.''  

%%%%%%%%%%%%%%%%%%%%%%%%%%%%%%%%%%%%%%%%%%%
\section{Summary} \label{sec:discuss}
%%%%%%%%%%%%%%%%%%%%%%%%%%%%%%%%%%%%%%%%%%%

We analyzed a one-point PDF for the peculiar velocity and the density
fluctuation. Here we start from the Gaussian distribution and analyze
the occurrence of non-Gaussianity. For the dust model,
in Eulerian linear approximation,
because the growth of the fluctuation is always linear, the distribution
always remains Gaussian. In Lagrangian linear approximation,
although the PDF of the density fluctuation becomes non-Gaussian,
the PDF of the peculiar velocity remains Gaussian, because
the peculiar velocity is proportional to the Lagrangian displacement.
Therefore the non-Gaussianity is produced by nonlinear effect.

Next we consider the effect of the pressure. If the pressure
exerts an effect,
the evolution of the fluctuation depends on the effect's scale.
Therefore even if we take only the Eulerian linear approximation,
the pressure may produce non-Gaussianity of the density fluctuation. 
However, from our results,
the PDF of the density fluctuation in the Eulerian linear
approximation is almost Gaussian during evolution.
In the same way, the PDF of the peculiar velocity is almost
Gaussian during evolution. Therefore we can conclude that
most of the non-Gaussianity of the PDF is produced
by the nonlinear effect of the perturbation.

We compared the evolution of the non-Gaussianity in the Lagrangian
linear approximation. As we showed in Figures~\ref{fig:P-delta-m3}
and \ref{fig:P-delta-m4}, the evolution depends on the 
``equation of state.'' In analyses of the PDFs of 
both the peculiar velocity and the density fluctuation, we can
distinguish the polytropic index. According to past
analyses~\cite{Morita01, Tatekawa02}, the behavior of the linear
perturbative solutions differ widely between the cases
of $\gamma=4/3$ and $\gamma=5/3$. In the case of $\gamma=4/3$,
the behavior of the solutions strongly depends on the scale
of the fluctuation. If we consider large-scale fluctuation,
because self-gravity dominates, the fluctuation grows.
On the other hand, if we consider small-scale fluctuation,
because the pressure dominates, the fluctuation oscillates
and decays. In the case of $\gamma=5/3$, the pressure affects
only the early stage. At the late time, the behavior of the perturbative
solution approaches that of ZA solutions. Therefore the
statistical quantities we computed, i.e., the variance,
the skewness and the kurtosis of the peculiar velocity
and the density fluctuation approaches to those in
the case of ZA~(Figures~\ref{fig:P-sigma}, \ref{fig:P-V-m3},
\ref{fig:P-V-m4}, \ref{fig:P-delta-m3} and \ref{fig:P-delta-m4}).

Although our result can apply until quasi-nonlinear stage is reached,
we can expect that an ``equation of state'' can be distinguished from
the growth of the non-Gaussianity.
If we can observe the PDF
of the density fluctuation in high-z region, i.e.,
the quasi-nonlinear region, we can find
the constraint of the character of the dark matter~\cite{DarkMatter}.

For comparing between the theoretical models and the observation,
spacial two-point correlation function
\cite{Totsuji69,Peebles74,CfA,LCRS,ESP,2dF,SDSS} can
also be important. We will
compute the correlation function and discuss the difference
between the theoretical models.

\ack
We are grateful to Kei-ichi Maeda for his continuous encouragement.
We thank Shuntaro Mizuno for useful discussion.
This work was supported by the Grant-in-Aid for Scientific
Research Fund of the Ministry of Education, Culture, Sports, Science
and Technology, Japan (Young Scientists (B) 16740152).


\clearpage

\section*{References}
\begin{thebibliography}{99}

\bibitem{Paddy93book}
Padmanabhan T 1993 {\it Structure Formation in the Universe}
(Cambridge: Cambridge University Press)

\bibitem{coles}
Coles P and Lucchin F 1995 {\it Cosmology: The Origin and Evolution
 of Cosmic Structure} (Chichester: John Wiley and Sons)

\bibitem{Peacock}
Peacock J 1999 {\it Cosmological Physics}
 (Cambridge: Cambridge University Press)

\bibitem{zel}
Zel'dovich Ya B 1970 {\it Astron. Astrophys.} {\bf 5} 84

\bibitem{Arnold82}
Arnol'd V I, Shandarin S F and Zel'dovich Ya B 1982
{\it Geophys. Astrophys. Fluid Dynamics} {\bf 20} 111

\bibitem{Shandarin89}
Shandarin S F and Zel'dovich Ya B 1989 \RMP
{\bf 61} 185

\bibitem{buchert89}
Buchert T 1989 {\it Astron. Astrophys.} {\bf 223} 9

\bibitem{saco}
Sahni V and Coles P 1995 {\it Phys. Rep.} {\bf 262} 1

\bibitem{Jones}
Jones B J T, Mart\'{i}nez V J, Saar E and Trimble V
2004 {\it Rev. Mod. Phys.} {\bf 76}, 1211

\bibitem{Tatekawa05}
Tatekawa T 2004 {\it Preprint} astro-ph/0412025

\bibitem{Paddy05}
Padmanabhan T 2005 {\it Preprint} gr-gc/0503107

\bibitem{munshi}
Munshi D, Sahni V and Starobinsky A A 1994
{\it Astrophys. J.} {\bf 436} 517

\bibitem{sahsha}
Sahni V and Shandarin S F 1996 {\it Mon. Not. R. Astron. Soc.}
{\bf 282} 641

\bibitem{Yoshisato98}
Yoshisato A, Matsubara T and Morikawa M 1998
{\it Astrophys. J.} {\bf 498} 48

\bibitem{gurbatov}
Gurbatov S N, Saichev A I and Shandarin S F 1989
{\it Mon. Not. R. Astron. Soc.} {\bf 236} 385

\bibitem{cms}
Coles P, Melott A L and Shandarin S F 1993
{\it Mon. Not. R. Astron. Soc.} {\bf 260} 765

\bibitem{mps}
Melott A L, Pellman T F and Shandarin S F 1994
{\it Mon. Not. R. Astron. Soc.} {\bf 269} 626

\bibitem{BT}
Binney J and Tremaine S 1987 {\it Galactic Dynamics}
 (Princeton: Princeton University Press)

\bibitem{budo}
Buchert T and Dom\'{\i}nguez A 1998
{\it Astron. Astrophys.} {\bf 335} 395

\bibitem{bdp}
Buchert T, Dom\'{\i}nguez A and P\'{e}rez-Mercader J 1999
{\it Astron. Astrophys.} {\bf 349} 343

\bibitem{domi00}
Dom\'{\i}nguez A 2000 \PR {\bf D62} 103501

\bibitem{domi02}
Dom\'{\i}nguez A 2002 {\it Mon. Not. R. Astron. Soc.}
{\bf 334} 435

\bibitem{Buchert05}
Buchert T and Dom\'{\i}nguez A 2005 {\it Preprint} astro-ph/0502318

\bibitem{adler}
Adler S and Buchert T 1999 {\it Astron. Astrophys.}
{\bf 343} 317

\bibitem{Morita01}
Morita M and Tatekawa T 2001 {\it Mon. Not. R. Astron. Soc.}
{\bf 328} 815

\bibitem{Tatekawa02}
Tatekawa T, Suda M, Maeda K, Morita M and Anzai H 2002 \PR
{\bf D66} 064014

\bibitem{Tatekawa05A}
Tatekawa T 2005 \PR {\bf D71} 044024

\bibitem{Kofman94}
Kofman L, Bertschinger E, Gelb J M, Nusser A and
 Dekel A 1994 {\it Astrophys. J.} {\bf 420} 44

\bibitem{Weinberg}
Weinberg S 1972 {\it Gravitation and Cosmology} (New York:
John Wiley and Sons)

\bibitem{Peebles80}
Peebles P J E 1980 {\it The Large-Scale Structure of the Universe}
 (Princeton: Princeton University Press) p 151

\bibitem{Paddy93}
Padmanabhan T and Subramanian K 1993 {\it Astrophys. J.}
{\bf 410} 482

\bibitem{Kayo01}
Kayo I, Taruya A, and Suto Y 2001 {\it Astrophys. J.}
{\bf 561} 22

\bibitem{COSMICS}
Ma C P and Bertschinger E 1995 {\it Astrophys. J.} {\bf 455} 7

\bibitem{Tatekawa04A}
Tatekawa T 2004 \PR {\bf D69} 084020

\bibitem{DarkMatter}
Ostriker J P and Steinhardt P 2003 {\it Science}
{\bf 300} 1909

\bibitem{Totsuji69}
Totsuji H and Kihara T 1969 {\it Pub. Astron. Soc. Japan}
{\bf 21} 211

\bibitem{Peebles74}
Peebles P J E 1974 {\it Astron. Astrophys.} {\bf 32} 197

\bibitem{CfA}
Geller M J and Huchra J P 1989 {\it Science} {\bf 246} 897

\bibitem{LCRS}
Jing Y P, Mo H J and B\"{o}rner G 1998 {\it Astrophys J.}
{\bf 494} 1

\bibitem{ESP}
Guzzo L {\it et al.} 2000 {\it Astron. Astrophys.}
{\bf 355} 1

\bibitem{2dF}
Hawkins E {\it et al.} 2003 {\it Mon. Not. R. Astron. Soc.}
{\bf 346} 78

\bibitem{SDSS}
Zehavi I {\it et al.} 2002 {\it Astrophys. J.} {\bf 571} 192

\end{thebibliography}
\end{document}